\documentstyle[prl,aps,epsfig]{revtex}
\def\beq{\begin{equation}}
\def\eeq{\end{equation}}
\def\bi{\begin{itemize}}
\def\ei{\end{itemize}}
\def\beqar{\begin{eqnarray}}
\def\eeqar{\end{eqnarray}}
\newcommand{\dx}{{ d^2 x}}

\newcommand{\dmu}{\partial_{\mu}}

\newcommand{\vev}[1]{\left< #1 \right>}
\newcommand{\cL}{{\cal L}}

\newcommand{\tlam}{\lambda}
\newcommand{\tg}{g}
\newcommand{\nn}{\nonumber \\ }

\newcommand{\Tr}{\hbox{Tr}}
\begin{document}
\textheight=23.6cm
\twocolumn[\hsize\textwidth\columnwidth\hsize\csname
@twocolumnfalse\endcsname

\title{A Step Beyond the Bounce:  Bubble Dynamics in Quantum Phase Transitions}
\author{Yoav Bergner and Lu\'{\i}s M. A. Bettencourt}
\address{Center for Theoretical Physics, Massachusetts Institute of
Technology, Cambridge MA 02139}

\date{\today}

\maketitle

\begin{abstract}
We study the dynamical evolution of a phase interface or bubble in the context
of a $\lambda \phi^4 + g \phi^6$ scalar quantum field theory. We use a 
self-consistent mean-field approximation derived from a 2PI effective action
to construct an initial value problem for the expectation value of the quantum 
field and two-point function. We solve the equations of motion numerically 
in (1+1)-dimensions and compare the results to the purely classical
evolution.  We find that the quantum fluctuations dress the classical
profile, affecting both the early time expansion of the bubble and the
behavior upon collision with a neighboring interface.  

\end{abstract}

\pacs{PACS Numbers: 11.27.+d, 64.60.Ht, 05.45.-a, 11.10.-z  \hfill MIT--CTP-3273}

%
%
\vskip2pc]

\section{Introduction}
\label{secI}

During first order phase transitions, bubbles or domains of the lower free energy phase (true 
vacuum) are nucleated in a metastable, or false vacuum, phase. 
Even at zero temperature, bubbles are induced by quantum effects, but they may also be
thermally activated.  The theory of droplet formation, describing the onset of 
nucleation, is by now venerably old and well established. It stretches back 
to the work of Becker and D\"oring \cite{BD} and Langer \cite{Langer} 
in statistical physics and later to that of Coleman \cite{Coleman} in the context of relativistic 
quantum fields, among others \cite{Review}. 

The theory of droplet nucleation, although successful, leaves almost all 
dynamical questions unanswered: what happens to the system once the bubble is nucleated?
The general phenomenology of bubble expansion and coalescence must
address how a semi-classical field solution (the bubble) propagates  
in the presence of quantum or thermal fluctuations for long times,
i.e. how these fluctuations interact quantum 
mechanically with the interface, and how the full self-consistent system may be described 
classically by hydrodynamics, for example of front propagation in media.

All of these questions can be easily posed and are, in principle, answerable 
in the context of quantum field theory. Tackling them quantitatively however requires 
a combination of non-perturbative analytical and numerical techniques that 
are just now beginning to emerge. 
The aim of the present paper is  to take the first steps toward studying the nucleation and 
dynamical propagation of bubbles together with their self-consistent quantum fluctuations  
in relativistic quantum field theory. 

The theory of droplet nucleation tells us that there are subcritical 
bubbles which decay away and also supercritical bubbles which feed on the energy
released by the phase transition to grow until the  true
vacuum phase has obliterated the false vacuum entirely. Coleman dubbed 
this the fate of the false vacuum~\cite{Coleman}.  

Details of the dynamics described heuristically above are notably
absent.  To address the question of what is the critical bubble size, what
is the shape or profile of the bubble as it expands, and whether  the
bubble wall experiences viscous drag, we need a thorough
understanding of the nonequilibrium quantum field dynamics.  A
classical analysis based on global properties of Lorentz invariance is
present in~\cite{Coleman}, but this of course leaves out fluctuations and
hence both virtual or real particles. 

The inclusion of (self-consistent) particles or fluctuations
leads to a panoply of new phenomena that must be considered for the complete description 
of the phase transition. Recognizing this, Coleman
 left a number of open questions concerning the effects of
fluctuations (particles) on interfaces and vice-versa \cite{Coleman}.
These issues are not manifest in the strict context of the nucleation problem.  

The first of Coleman's questions is,  what happens when a bubble encounters 
particles? This phenomenon is central to scenarios of early
Universe baryogenesis and has been addressed in this context to some extent
\cite{Baryogenesis}.  Baryogenesis remains the most important
motivation for the study of bubble wall dynamics in  relativistic
settings. Several works \cite{Moore,Laine} have recently addressed the problem of computing the 
asymptotic velocity and shape of Higgs field bubbles at temperatures near 
the electroweak phase transition. These approaches treat the bubble wall 
as a classical field background immersed in a bath of thermal fluctuations 
which obey effective transport equations for their occupation number distributions.
This treatment  is appropriate if the bubble wall moves sufficiently
slowly, contains only ``soft gradients'',  and if quantum coherence is unimportant.
Thus a transport approach will necessarily fail at sufficiently low temperatures 
and/or under severe supercooling. In these more difficult cases, the direct field theoretical 
methods developed here become essential.
Quantum first order phase transitions in non-relativistic systems \cite{QFOPT} may  provide an
interesting laboratory for testing the non-relativistic analogue of
the zero-temperature methods
described below.  

Coleman's other questions are concerned with the possibility that bubbles  may be induced 
by fluctuations (and perhaps even created at particle scattering experiments 
\cite{Rubakov})  and with particle production resulting from the collision of two bubble walls. Both phenomena necessitate a dynamical non-perturbative treatment of quantum field 
theory valid for long times. For this reason they have remained poorly understood.

In recent years, the availability of numerical methods to solve for
the time evolution of quantum fields has given rise to a resurgence of
interest in such problems.  The causal formalism suited to initial
value formulations of field theory dynamics has been employed in
various approximation schemes in an effort to isolate the relevant
features of a quantum kinetic theory from first
principles~\cite{schwinger,cjt,calhu,coopmott,berges}.

In this paper we consider a scalar quantum field theory which exhibits a
first-order phase transition. Assuming that the field is in the ``false vacuum''  before the
transition and is brought out of equilibrium by the nucleation of
bubbles in this ``true vacuum'' phase, we study the detailed dynamics of bubbles
which we impose as initial conditions.  Because of the computational
effort required in the quantum theory, we restrict our attention to
(1+1)-dimensional spacetime.
 
We consider the purely classical field evolution as well as a self-consistent
quantum evolution in the Hartree approximation at zero temperature.
The generalization of the formalism to include both higher-order
interactions~\cite{Aarts} and/or finite temperature 
is straightforward.   Thermal effects lead to qualitatively different
physics, and we intend to analyze these physical consequences in
detail in a future work.

The main results of this paper are: 1)
Whether a true vacuum bubble is critical is determined by the extremization of
the energy, not the action.  While this should be obvious from the
point of view of an initial value problem, there has traditionally
been some confusion of the critical (spacetime) radius for the bounce,
$R_B$ with the critical (purely spatial) radius for growth, which we
label $R_E$.  The two values are related by a constant
of proportionality 
$${R_E \over R_B} = {(d-1) \over d}$$ 
in $d$ spatial dimensions; hence for (1+1)-dimensions, any
bubble is critical.  More precisely, the critical bubble size is
constrained in one dimension only by the thickness of the bubble. 
2) The bounce determines the correct profile of the bubble wall, but induced,
   super-critical bubbles with larger or smaller radii still grow and
   asymptote to shifted light-cones.  The bounce solution is unique and
identifiable in that it asymptotes to the light cone from the origin.
These results are already manifest in the classical description. 
3) Including quantum effects at the level of the Hartree approximation
   does not change the qualitative features--constrained by Lorentz 
invariance--of bubble growth at
   zero-temperature.  However, in much the same sense that quantum
fluctuations render the quantum effective potential different from the
classical one, they do affect the detailed shape of the bounce.  Hence,
4) The proper description of quantum bubble dynamics necessitates a
   self-consistent bounce which includes a prescription of the quantum
   fluctuations at the time of nucleation.
5) The behavior of colliding bubbles does indicate a qualitative
   difference between the classical and quantum behavior.  In our
model, the
   classical bounce appears remarkably stable against bubble
   coalescence--exhibiting elastic collisions off neighboring
   expanding bubbles for very long times (possibly forever).  The quantum evolution, on
the other hand, displays the more expected
   behavior wherein the bubbles disappear by transferring energy
to intermediate frequencies on a time scale of the same order of magnitude as the bubble size.

In Section~II we summarize the semiclassical
theory and its predictions;  we present our model and highlight some
of the classical dynamical details which emerge anew due to the
consideration of the dynamics as an initial value problem and due to
our specialization to (1+1)-dimensions.  In Section~III, we
extend the analysis of the dynamics to a self-consistent 
Hartree-like approximation and discuss the simplifications it involves.  
We summarize our results for the propagation and collision 
of bubble walls in the quantum theory in Section~IV. In Section~V we discuss 
the many interesting possibilities for the application of the methods of this paper 
to related questions as well as the refinements necessary to render the 
long-time evolution of self-consistent quantum fluctuations more realistic.

\section{(semi)classical dynamics}
\label{secII}

Ultimately, we would like to understand the quantum dynamical evolution of a
generic bubble of true vacuum (induced perhaps by coupling to other fields or
sources).  We should naturally do first what we can in the classical
regime, where we may apply the literature on
semiclassical field theory methods in the bubble nucleation
problem~\cite{Langer,Coleman} and connect to other numerical 
studies~\cite{Ferrera}.
The relativistic picture was
elegantly framed by Coleman in Ref.~\cite{Coleman}, so we shall parallel
that analysis, working out a specific example in full detail.     

Coleman set out to compute the decay rate of the false vacuum in a
scalar theory described by the Lagrangian
\beq
\cL = \frac12 (\dmu \phi)^2 - V(\phi) \  . 
\eeq
By analogy with the semiclassical analysis of barrier penetration, he
obtained the exponent in the vacuum decay rate in terms of an
instanton solution of Euclidean spacetime which he called the bounce.  
The bounce function is a saddle point of the Euclidean action, 
\begin{eqnarray}
\label{action}
S_E[\phi] = \int d^dx~d\tau \left\{ \frac12 \left( 
{d\phi  \over d\tau} \right)^2 + \frac12 \left( 
{d\phi  \over d \vec{x}} \right)^2 + V(\phi) \right\} ,
\end{eqnarray}
hence it satisfies Euclidean ``equations of motion''.  The solution is
subject to appropriate boundary conditions at the origin and at
infinity.  It is understood that such a function will always exist and
will be $O(D)$ invariant for $D=d+1$ spacetime dimensions, thus depending
only on a radial coordinate $\rho=\sqrt{\tau^2+x^2}$, $\phi_b = \phi_b(\rho)$.
Then the bounce equation and corresponding 
boundary conditions take the form
\beqar
&& {d^2 \phi_b \over d \rho^2} + {D-1 \over \rho} {d \phi_b \over d \rho } = 
{\delta V(\phi_b) \over \delta \phi_b}, \\
&& \lim_{\rho \rightarrow \infty} \phi(\rho)= \phi_+ \qquad \mbox{and} \qquad
\left.\frac{d\phi}{d\rho}\right|_{\rho=0} =0 \ .
\label{boundary}
\eeqar
A happy consequence of the bounce solution is that 
the real-time classical equation of motion is just the analytic
continuation of the bounce equation to real time and thus is solved by the
analytic continuation of the bounce from Euclidean spacetime to
Minkowski, i.e. with $\tau \rightarrow i t$,
\beq
\phi_b(\rho) \rightarrow \phi_b(\sqrt{x^2-t^2}) .
\eeq
The shape of the bounce becomes the profile of the bubble
wall, and this shape remains unaltered as the wall describes a
hyperbola in spacetime, reaching the speed of light asymptotically.  
Lorentz invariance allows us to solve for the bubble profile 
in all space and time and not as an initial value problem.

A number of features of the spontaneous decay of the false vacuum are
severely constrained. The most favorable shape of the bubble is 
constrained by stationarization of the action. 
That its asymptotic velocity is $c$ (or 1 in natural units) 
is enforced by $O(D)$ invariance of the bounce in
Euclidean spacetime or $O(D-1,1)$ Lorentz invariance.  Thus it is
impossible in the absence of Lorentz violating fluctuations for the
bubble wall to experience drag. Classically, any drag would have to
be put in by hand in the equations of motion\footnote{We have 
verified that the addition of a simple drag term $\eta \partial_t \phi$ 
in the dynamical field equations does indeed result in an asymptotic 
interface velocity smaller that that of light. 
The study of the stochastic bubble wall trajectory in the presence of phenomenological 
damping and noise is in itself an interesting problem, see \cite{stochastic}.}.

\begin{figure}[h]
\begin{center}
\psfig{file=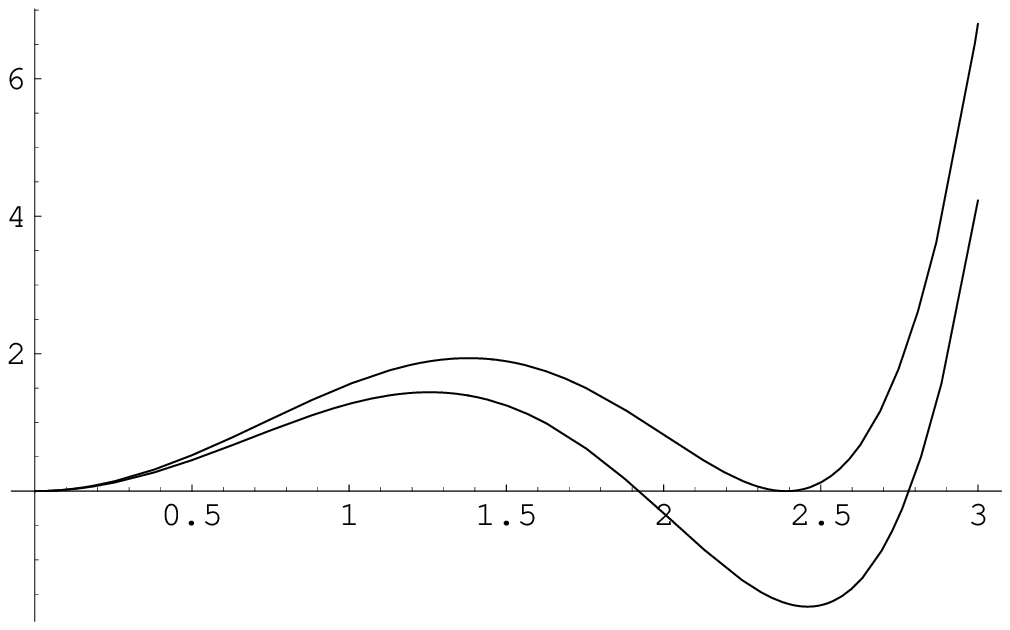,width=3in,height=2in,silent=}
\end{center}
\caption{The classical potential (lower curve) for the bounce calculation with
minima at $\phi_-=0$ and $\phi_+=2.46$. The upper curve is the
degenerate potential (see Eq.~(\ref{potential2})), used to compute in
the thin wall approximation.  Its second minimum occurs at $\phi=2.39$.}
\label{fig1}
\begin{picture}(1,1)(50,240)
\put(230,430){$\gamma=0$}
\put(270,330){$\phi$}
\put(30,450){$V(\phi)$}
\end{picture}
\end{figure}

Let us make the analysis concrete by specializing to a potential
of the form
\begin{eqnarray}
\label{potential1}
V[\phi] = {1 \over 2 } m^2 \phi^2 + \lambda \phi^4 +g \phi^6 .
\end{eqnarray}
For the purposes of comparison with an analytic, thin-wall
approximation, it is convenient to have a potential with degenerate
minima in a simple parametric limit.  We can rewrite the potential in
terms of the parameters $\phi_0^2$ and $\gamma$ such that  
\begin{eqnarray}
V[\phi] = g \phi^2 \left( \phi^2 - \phi_0^2 \right)^2 - \gamma \phi^2.
\label{potential2}
\end{eqnarray}
where $\phi_0^2 = -\lambda/2g$ and $\gamma=(\lambda^2/g-2 m^2)/4$. 
The thin wall approximation relies on neglecting the term proportional 
to $\gamma$. In practice the thin wall obtains in the limit where the energy 
difference between the two minima $\epsilon=V(\phi_-)-V(\phi_+)=\gamma \phi_0^2$ vanishes. 
As a working example we use the parameters $m^2=4, \lambda=-0.8$ 
and $g=0.07$. Then $\gamma=0.29$.
With these choices the potential for the bounce calculation is shown 
in Fig.~\ref{fig1}. Also shown is the degenerate potential ($\gamma=0$).

\begin{figure}
\begin{center}
\psfig{file=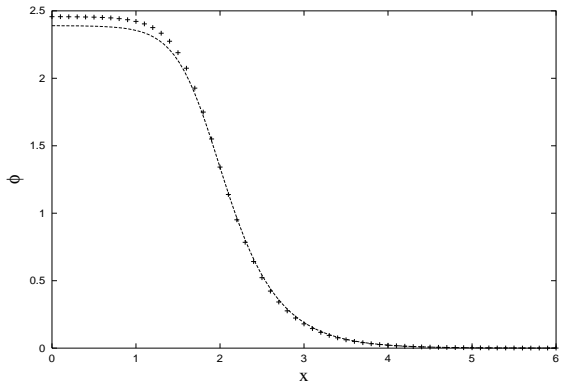,width=3in,height=2in,silent=,angle=0}
\end{center}
\caption{The numerical solution for the bounce profile (points) and the 
thin wall approximation from Ref.~\protect\cite{Brasil}. The thin wall
approximation agrees well with the exact numerical computation where
it should, i.e. at $\rho \simeq R_c=1.82$.}  
\label{fig2}
\end{figure}

We construct the bounce solution in the thin wall approximation,
following Ref.~\cite{Coleman}, and by direct numerical integration.  
The thin wall bounce consists of solving for the 
configuration of $\phi$, piecewise in $\rho$ such that
\begin{eqnarray}
&& \phi = \phi_-=\phi_0 + O(\gamma), \qquad \rho \ll R \nonumber \\
&& \phi = \phi_{\rm wall}(\rho-R), \qquad \rho \simeq R   \\
&& \phi = \phi_+=0, \qquad \rho \gg R. \nonumber 
\end{eqnarray}
Here $\phi_{\rm wall}$ is computed in the degenerate potential,
i.e. by solving  
\begin{eqnarray}
&& {d^2 \phi_{\rm wall} \over d \rho^2} = U'[\phi_{\rm wall}]; 
\label{thinwall} \\ 
&& U[\phi] = g \phi^2 \left( \phi^2 +{\lambda \over 2 g} \right)^2.
\label{potentialU}
\end{eqnarray}
The solution has been obtained previously~\cite{Brasil}:
\begin{eqnarray}
\label{thinbubble}
\phi^2_{\rm wall} (\rho) = {\phi^2_0 \over 1 + e^{\mu \rho} },
\end{eqnarray}
where $\mu=\sqrt{8g}\phi_0^2=4.21$, and $\mu^2$ is the second derivative
of the potential $U''[\phi_0]$ evaluated at $\phi_0$.  Up to 
a correction of order $\epsilon$, it is the mass of excitations around
the true minimum.  
To complete the thin wall approximation $R$ is determined variationally,
as the value that extremizes the Euclidean action. In the thin wall 
approximation to the action, the interface itself amounts to a surface term, 
whereas the contribution from the two piecewise constant parts is proportional 
to the bubble's spacetime volume. Hence in (1+1)-dimensions
\begin{eqnarray}
\label{thinwallR}
S_E = - \pi R^2 \epsilon + 2 \pi R \sigma,
\end{eqnarray} 
where $\sigma$ is a surface tension. 
The stationary point of (\ref{thinwallR}) is equivalently
the zero-energy value of $R$ in configuration space:
\beq
\label{energy}
-R \epsilon + \sigma = 0 \ ,
\eeq
so that $R=\sigma/\epsilon$ with  
\begin{eqnarray}
\sigma = \int d x \left[ {1 \over 2} \left( 
{d \phi_{\rm wall} \over d x} \right)^2 + U[\phi_{\rm wall}] \right] \ .
\end{eqnarray}
With our parameters, $\sigma=3.05$, and $R=1.82$. 
The two solutions, the exact bounce computed numerically and in the 
thin wall approximation, are shown in Fig.~\ref{fig2}. 

We note that the thin wall bounce is a good approximation to the exact 
solution in the neighborhood of the inflection point i.e. at 
$\rho \simeq R_c=1.82$.  The true condition for the validity of the 
approximation~\cite{Coleman} is that $\mu R \gg 1$. With our parameters, 
$\mu = \sqrt{8 g} \phi_0^2=4.21$, so that $\mu R = 7.66 \gg 1$, which 
is shy of an order of magnitude larger than unity.  In fact, as depicted 
in Fig.~\ref{fig2}, the bubble wall does not appear very ``thin''. 

\begin{figure}
\begin{center}
\psfig{file=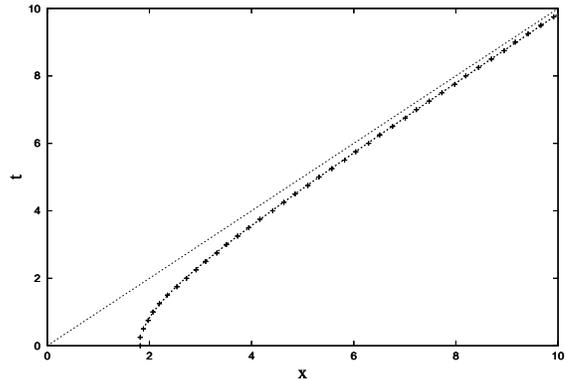,width=3in,height=2in,silent=,angle=0}
\end{center}
\caption{Spacetime trajectory for a (1+1)-dimensional 
bubble started at rest at $t=0$ is well described by the  hyperbola $x
= \sqrt{R^2 +t^2}$ with $R=1.82$.}
\label{bouncetrajectory}
\end{figure}

To confirm the predictions of the Euclidean solution and test 
our numerical methods we now solve the classical real time equations of 
motion (i.e. in Minkowksi space) for the field $\phi$ explicitly.  
As initial conditions, we use a bubble at rest with the profile given by
the approximate analytic form (\ref{thinbubble}).  By following one
point on the bubble wall with time, we observe the predicted
trajectory, shown in Fig.~\ref{bouncetrajectory}.

We have so far said nothing about induced vacuum decay, in which a
bubble may be nucleated with an arbitrary size and shape.    
As seen in Eq. (\ref{energy}), the spontaneous decay of the vacuum via the bounce  
costs exactly no energy. However, if the energy of an
initial bubble configuration may be  accounted for from another source, 
it need not be zero.  We would like to know how such a 
bubble will evolve given what we know about the bounce. 
We first consider varying the size of the bubble without changing its
interface profile. 

\begin{figure}
\begin{center}
\epsfig{file=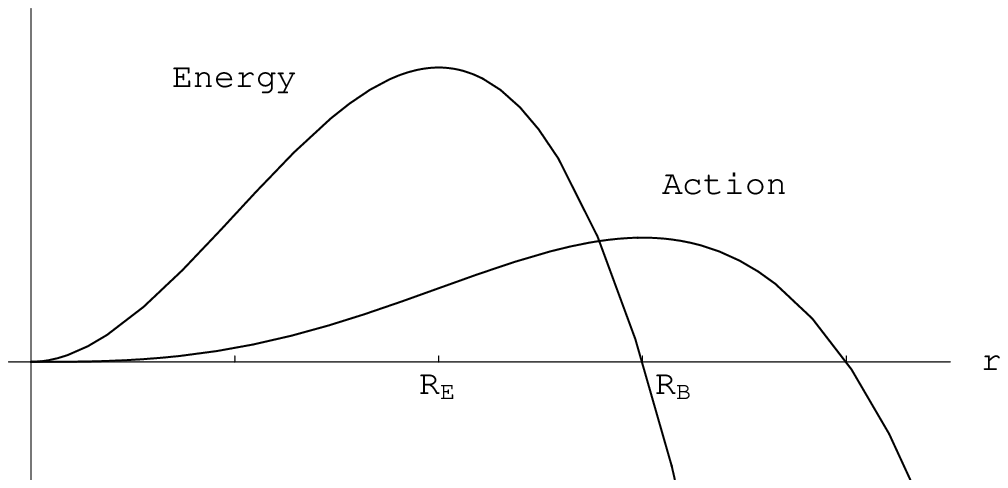,width=2.75in,height=1.25in,silent=,angle=0}
\end{center}
\begin{center}
(a) $d=3$
\end{center}
\vspace{-0.325in}
\begin{center}
\epsfig{file=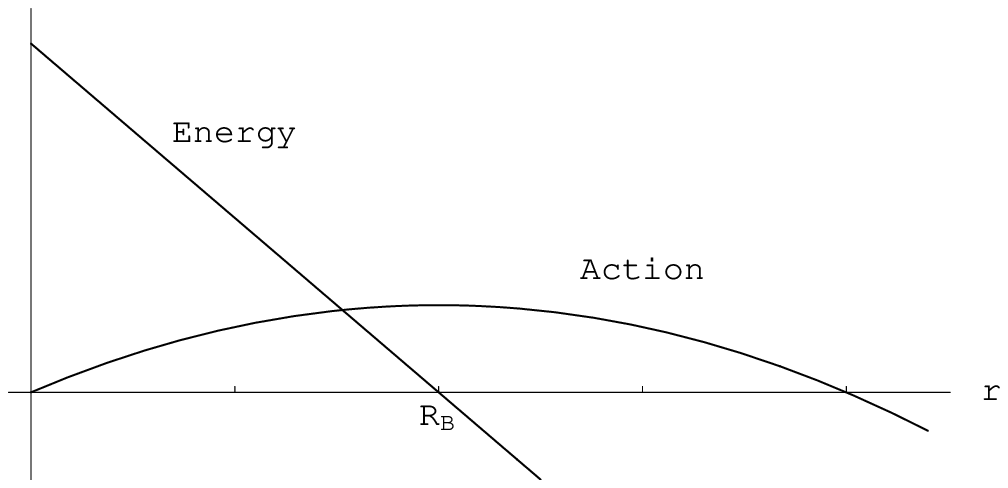,width=2.75in,height=1.25in,silent=,angle=0}
\end{center}
\begin{center}
(b) $d=1$
\end{center}
\caption{The dependence of energy and action on the radial coordinate
in the thin wall approximation: (a) a ``spherical'' bubble
in 3d. The critical radius for growth $R_E$ is the maximum
of the energy and is smaller by a factor of $(d-1)/d$ than the radius $R_B$ which
maximizes that action and corresponds to the zero of energy. (b)
In 1d, the critical radius $R_E$ goes to zero.}
\label{3D1Denergy}
\end{figure}

Generally the fate of an induced bubble--growth or decay--can be determined 
by considering the energetics of the corresponding initial value
problem. 
In this case, we do not have a solution for all spacetime, but only a
spatial profile at some initial time $t=0$.     
We will assume that the bubble wall is initially at rest, i.e. that its
kinetic  
energy is zero \footnote{This is a weak assumption from the point of view of the initial value problem.  From the Euclidean spacetime analysis, it is a consequence, not an assumption}. The total energy is given by
\begin{eqnarray}
E = \int d^d x \ {1 \over 2} \left( \partial_t \phi_{\rm b} \right)^2 
+ {1 \over 2} \left( \nabla \phi_{\rm b} \right)^2 + V(\phi_{\rm b}) ,
\end{eqnarray} 
where the static part can again be computed in the thin wall limit. 
It is $E_{\rm static} = \sigma S_d - \epsilon V_d$, where $S_d$ and 
$V_d$ are the surface and volume of the bubble in $d=D-1$ {\it spatial} 
dimensions. The energy has a local maximum at $R_E= (d-1) \sigma/\epsilon$  
(recall that the bounce has an action extremum at $R_B=d\sigma/\epsilon$). 
For the interface to move while globally conserving energy it is necessary 
that it can lower its static energy, so that the difference is converted into 
kinetic energy. Bubbles with $R>R_E$ do so by growing while those with 
$R<R_E$ must shrink.  Hence $R_E$ defines the critical radius for growth. A
particular case is that of the bounce which corresponds  
to the choice of $R_B$ which enforces $E_{\rm static}=0$. 
The bounce {\it always} falls in the class $R>R_E$ and therefore
always grows.  However, it is not the case, as is frequently assumed
in the literature, that the bounce radius is the critical radius as
defined by the onset of dynamical growth.  We plot an example of the energy in
3~spatial dimensions in Fig.~\ref{3D1Denergy}(a).  

The case of (1+1)-dimensions ($d=1$ above) is special because the interface 
energy saturates:  once the bubble is larger than twice the thickness of
the wall, the ``surface'' term no longer depends on the size of the bubble.  
Since the  bubble can gain kinetic energy by converting false vacuum, 
a thin wall bubble will {\it always  grow} in (1+1)-dimensions, 
i.e. $R_E=0$.  This is seen clearly in Fig.~\ref{3D1Denergy}(b), where 
the thin wall approximation is again assumed.  If $\mu$ is a measure of 
the inverse thickness of the bubble, then the thin wall approximation 
amounts to $\mu R \gg 1$.  In practice, the finite size of the wall profile constrains $\mu$ 
so that $R_c\simeq 1/\mu$ gives a lower limit on the critical 
bubble size.

\begin{figure}
\begin{center}
\psfig{file=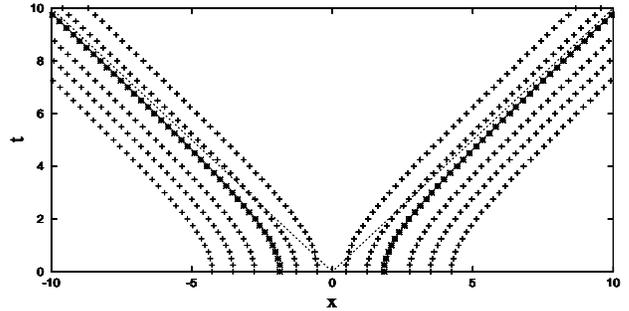,width=3.2in,height=1.6in,silent=,angle=0}
\end{center}
\caption{Spacetime trajectories for the motion of several (1+1)-dimensional 
interfaces started at rest at $t=0$ and with different bubble radii. All trajectories 
are well described by hyperbolae $x = \pm(x_0 + \sqrt{R^2 +t^2}$), with origins
set at $t=0$. Thus the interface velocity at large times approaches the speed 
of light as in the case of the bounce, where $x_0=0$.}
\label{trajectories}
\end{figure}

We verified this behavior numerically as well as the fact that the thin wall 
prediction of $R_E^{d=2}=R_B^{d=1}=1.817$ is an excellent 
approximation to the real critical value for bubble growth in $d=2$, 
which occurs between $R=1.810-1.815$.  Changing the radius amounts to
using $\phi(\rho-R)$ with the analytic form (\ref{thinbubble}) 
and variable $R$ as an initial condition.  We show the trajectories
obtained for bubbles with different initial sizes in
Fig.~\ref{trajectories}.  Note that the trajectories for bubbles with
radii smaller or larger than the bounce
asymptote to shifted light cones.  Naively, one might have expected
all solutions to approach the light cone from the origin.

\begin{figure}
\begin{center}
\psfig{file=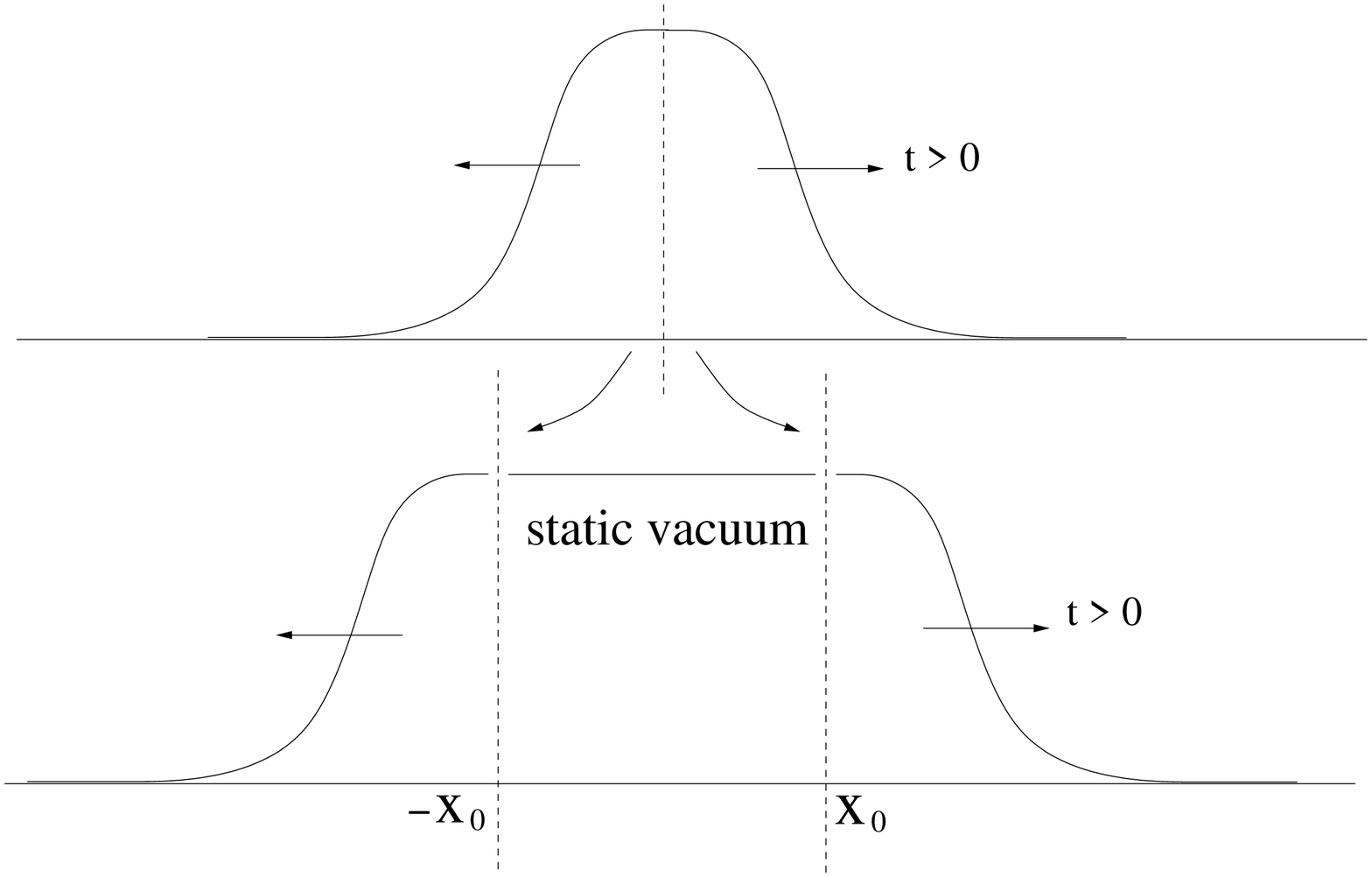,width=3in,height=1.75in,silent=,angle=0}
\end{center}
\caption{Schematic construction at $t=0$ of a solution to the real
time equations of motion for all $t>0$.}
\label{piecewise}
\end{figure}

In order to explain these trajectories, we explicitly construct a
piecewise solution for all (real) time with an arbitrarily initial
size.  The real time equations of motion are invariant under a
coordinate shift $x \rightarrow x+x_0$.  Consider cutting the function $\phi_b(x,t)$
along $x=0$ at $t=0$ and 
shifting the positive $x$ piece by a positive amount $x_0$ and the
negative $x$ piece by $-x_0$.  In the gap created in the center,
extend the value of $\phi(x=0)$ to be constant. This
function looks exactly like the bounce at the interfaces but has more
true vacuum sandwiched in the middle, see
Fig.~\ref{piecewise}.  The
boundary conditions on the bounce (\ref{boundary}) ensure
continuity up to and including first derivates at $\pm x_0$.   (Excising a homogeneous region in the center,
one can similarly generate a smaller bubble.)

A constant field in true vacuum is a static solution of the equations
of motion, hence the piecewise solution will be static in the region
$|x| < x_0$ and expand like the bounce outside this region.  Moreover, we recall that
although $R$ is determined variationally in the thin wall approximation--this freedom is exploited by our initial condition--solving
the bounce  equation directly (under the assumption of $O(D)$
symmetry) determines $R$ uniquely.  The coordinate shift transforms
\beq
\label{shift}
\phi(\sqrt{x^2-t^2}-R) \rightarrow \phi(\sqrt{(x\mp x_0)^2-t^2}-R) \ ,
\eeq
so the trajectory of the interface
is given  by  $x = \pm (x_0+\sqrt{R^2+t^2})$ in real 
time, implying that bubbles induced with different sizes will asymptote to 
different light cones as in Fig.~\ref{trajectories}.

So far we have altered the bubble size but not its profile.  Since to
distort the shape would be to perform a variation in function space, 
it is much more difficult to quantify the difference in the evolution of an 
arbitrary bubble from that of the bounce.  Heuristically, we understand 
that if a bubble starts out with the wrong profile, it will
try to change its shape to the bounce before it expands as described
above.  The critical  size argument can be affected by this: 
since a small bubble shrinks as it deforms, a slightly supercritical bubble 
with the wrong shape may still collapse away.  
A large initial bubble on the other hand will convert the false vacuum 
energy into kinetic modes which both expand and distort the bubble.
We have observed both types of behavior in numerical evolutions.

\section{Quantum bubbles}
\label{secIII}
Extending the analysis of the previous section to quantum fields
introduces a host of new challenges.  From the point of view of
dynamical equations of motion, we must now consider operator
equations, which in fact translate into an infinite Dyson-Schwinger  
hierarchy of equations for the $n$-point Green's functions of
the theory.  For practical purposes, the hierarchy must be truncated
somehow, and this truncation introduces an approximation, often in
the form of a self-consistent ansatz~\cite{SD}.
Furthermore, even upon keeping a finite number of connected correlation 
functions the resulting system of coupled equations still describes an
infinite number of degrees of freedom with non-linear
interactions. Such a system  
must in general be solved numerically in a finite computer, which can
lead to artifactual effects. 
Thus it is of utmost importance to identify practical approximation schemes 
that capture the qualitative essence of the dynamical evolution of a 
quantum field theory as compared with its classical counterpart.
This challenge is not devoid of uncertainty since not much is known about the 
evolution of truly quantum many body systems far away from thermal equilibrium.

The formulation of the problem can be cast into one formalism which
has been developed and explored in several works and has resurfaced in
recent years with renewed vigor~\cite{cjt,calhu,berges}.  It is based
on a two-particle irreducible (2PI) effective action for the field and 
the two-point function
$\Gamma[\phi,G]$ formalized by Cornwall, Jackiw and Tomboulis
(CJT)~\cite{cjt}.  The Schwinger-Keldysh closed time path (CTP) is employed
to make causality explicit through an appropriate real time
prescription in the path integral time contour and associated Green's
functions;  hence it is frequently referred to as the CJT or the 2PI-CTP
formalism.  It is in the evaluation of  $\Gamma[\phi,G]$ that
some sort of expansion (in loops or in powers of the coupling constant
or in powers of $1/N$ for example) is carried out.

In this paper we shall employ the 2PI-CTP formalism at the level of
the Hartree approximation, a mean-field approximation which
amounts to keeping the 2PI diagrams which are lowest order in coupling constants 
(bubble diagrams).  The Hartree approximation is well understood to be 
equivalent to a Gaussian variational ansatz in the Schr\"odinger 
functional formalism \cite{Schroedinger} and results
in Hamiltonian dynamics \cite{Hamiltonian}.  
It is also qualitatively similar to the
systematic large-$N$ approximation at leading order in $1/N$.  For our
purposes, $N=1$ and the Hartree approximation is motivated chiefly
because it is extremely simplifying.  A large $N$ scalar theory is also 
inappropriate for the description of first order transition dynamics 
as it results invariably in second order critical phenomena (unless 
the $O(N)$ symmetry is explicitly broken, whence criticality is erased 
and the transition becomes an analytical crossover). 

Time evolution of spatially inhomogeneous quantum fields, even at 
this level of approximation has only recently been produced 
numerically~\cite{smit,luis,braghin}.  To our knowledge, the growth of a
critical self-consistent quantum field bubble has not been previously 
demonstrated. 

The starting point for the 2PI-CTP formalism is the effective action
$\Gamma[\phi,G]$ of~\cite{cjt}:
\beqar
\Gamma[\phi,G]= && S[\phi] + \frac12 i \Tr \ln G^{-1} 
	+\frac12 i\Tr D^{-1}(\phi) G  \nn
&& + \Gamma_2 (\phi,G) + \hbox{const.} \ ,
\eeqar
where $i D^{-1}(\phi)$ is the classical inverse propagator
\beqar
i D^{-1}(x,y;\phi) = \frac{\delta S[\phi]}{\delta\phi(x)\delta\phi(y)}
\eeqar
and $\Gamma_2$ sums the 2PI vacuum to vacuum diagrams with
propagators set to $G$ and interaction vertices obtained from the
shifted Lagrangian $\cL[\phi \rightarrow \phi+\varphi]$.

\begin{figure}
\begin{center}
\leavevmode
\psfig{file=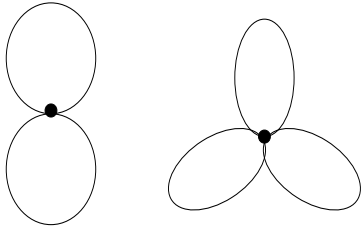,width=2in,height=1.33in,silent=,angle=0}
\end{center}
\caption{Vacuum bubble diagrams in the Hartree approximation for the 
2PI effective action of our model.  
Lines denote the full propagator $G$ in the nontrivial $\phi$ background.}
\label{diagrams}
\end{figure}

In the Hartree approximation, we consider 2PI diagrams in
Fig.~\ref{diagrams}, hence
\beqar
\Gamma_2 &=& -3(\tlam + 15\tg\phi^2)\int \dx \, G(x,x)^2 \nn
&&  \qquad \qquad \qquad -15 \tg \int \dx \, G(x,x)^3 \ . 
\eeqar

We observe that the two-point function enters into the Hartree 
approximation only as a local function evaluated at one spacetime point.  
This is apparent from the diagrams in Fig.~\ref{diagrams} which, upon
opening a line, become proper self-energy diagrams.  The inclusion of
any diagram with two or more vertices would introduce non-local
kernels in the equations of motion.  

The stationarity conditions on the effective action in the absence 
of sources, 
\beqar
\frac{\delta \Gamma[\phi,G]}{\delta \phi}=\frac{\delta
\Gamma[\phi,G]}{\delta G} = 0 ,
\eeqar
lead to equations of motion for the field
\beqar
\Bigl(\Box+m^2+4\tlam\phi^2+6\tg\phi^4+12(\tlam + 5\tg\phi^2) G(x,x) \label{fieldeq} \nn
\qquad \qquad \qquad + 90\tg \int \dx \, G(x,x)^2 \Bigr)\phi=0 \ , 
\eeqar
and for the Wightman two-point function
\beqar
\Bigl(\Box+m^2+12\tlam\phi^2+30\tg\phi^4+12(\tlam + 15\tg\phi^2) G(x,x) \label{propeq} \nn
\qquad \qquad \qquad \qquad \qquad + 90 \tg G(x,x)^2 \Bigr) G(x,x')=0 \ .
\eeqar 

Let us remark that there are logarithmic and quadratic divergences in
the bare theory in the self-energy and energy-momentum tensor respectively.
However, any scalar field theory in (1+1)-dimensions is renormalizable using the standard counterterm
procedure.  For the $\phi^6$ interaction, this has been carried out explicitly in an early analysis of the
quantum effective potential~\cite{Townsend}: the logarithmic divergence from the
scalar loop in (1+1)-dimensions affects both the mass and the quartic
coupling constant $\lambda$, which need to be renormalized.  In
practice, even though we solve the equations of motion 
numerically on a discretized spatial lattice, we introduce time-independent
counterterm subtractions as renormalization conditions for the ``physical''
mass, quartic coupling and energy density in vacuum.  We fix the
physical (dressed) mass to coincide with the bare mass of the classical theory,
i.e. we employ a subtraction (detailed below) which sets the two-point function to zero
in the false vacuum.  The quadratic divergence in the energy density
is just the usual zero-point energy of vacuum, and again we choose to
set the energy density in the false vacuum to zero. 

In order to solve the system of equations, we need to specify initial
data for the field and the two-point function.  This is more or less
straightforward for the field expectation value, which we identify 
with  the classical field bubble profiles of Section~\ref{secII}. 
Thus, as an {\em ansatz}, we initialize the quantum field 
expectation to have the same shape as the bounce function described 
in Eq.~(\ref{thinbubble}).  The initial
specification of the two-point function, however, is 
informed by the full details of the nonequilibrium ensemble.  As a
starting point, we use the most naive initial conditions possible,
which is to ignore the formation of the initial bubble completely.  In other
words, the two point function is initialized for a zero- or
finite-temperature distribution about the false vacuum everywhere in
space, including inside the bubble.  We are in the process of
improving this prescription and comment on this below.
 
We proceed by decomposing the equal-spacetime propagator in a mode basis.   From here on we use $x$ to denote the space coordinate alone instead of spacetime, i.e. $x^{\mu}=(x,t)$ and $G(x^{\mu},x^{\mu})$ can be written as
\beqar
G(x,t) = \sum_{k} |\psi_k(x,t)|^2(2 n_B(k)+1) \ ,
\label{G}
\eeqar
where $n_B(k)$ is the Bose-Einstein distribution
\beqar
n_B = \frac{1}{1-e^{\beta\omega_k}} \ .
\eeqar
The mode functions $\psi_k$ are initialized in a plane wave basis with the mass of 
excitations around the false vacuum:
\beqar
\psi_k(x) = \frac{e^{ik\cdot x}}{\sqrt{2 \omega_k}} \ ,
\eeqar
with
\beqar
\omega_k = \sqrt{k^2 + m^2} \ .
\eeqar

The mass in the dispersion relation is self-consistently
dependent on the values of the mean field and
fluctuations.  The log-divergent, zero-temperature part of $G$ is
renormalized by the subtraction of $G(x,t)$ evaluated at $x=t=0$,
\beqar
G_R(x,t) = G(x,t)-G(0,0) \ .
\eeqar
This is equivalent to mass and coupling constant renormalization.

Finally we evolve the field and two-point function according the
 equations (\ref{fieldeq}) and (\ref{propeq}) on a discrete
one-dimensional lattice.  We employ a variety of lattices corresponding
 to different physical volumes and lattice spacings and use a fourth
 order symplectic integrator to step in time.  The range in the
 number of lattice points was between 256 and 2048, with typical
 lattice spacing of 1/64 to 1/128.  Diagnostics performed at
regular time intervals give a series of cinematic snapshots of the
evolution and measure the components of the energy momentum
tensor $\vev{T_{\mu\nu}}$ as well as the separate components of the 
energy $\vev{T_{00}}$, i.e. potential, gradient and kinetic energy.

The results for the evolution of the quantum fields are shown in the 
next section, where they are also compared to the purely classical 
field bubbles of Section~\ref{secII}.

\section{Numerical results}
\label{secIV}

In Section~\ref{secII} we have already discussed the paramount importance 
of Lorentz symmetry in constraining many of the properties of bubble wall 
propagation. These constraints clearly also apply  in the 
quantum theory at zero temperature.  

\begin{figure}
\begin{center}
\leavevmode
\psfig{file=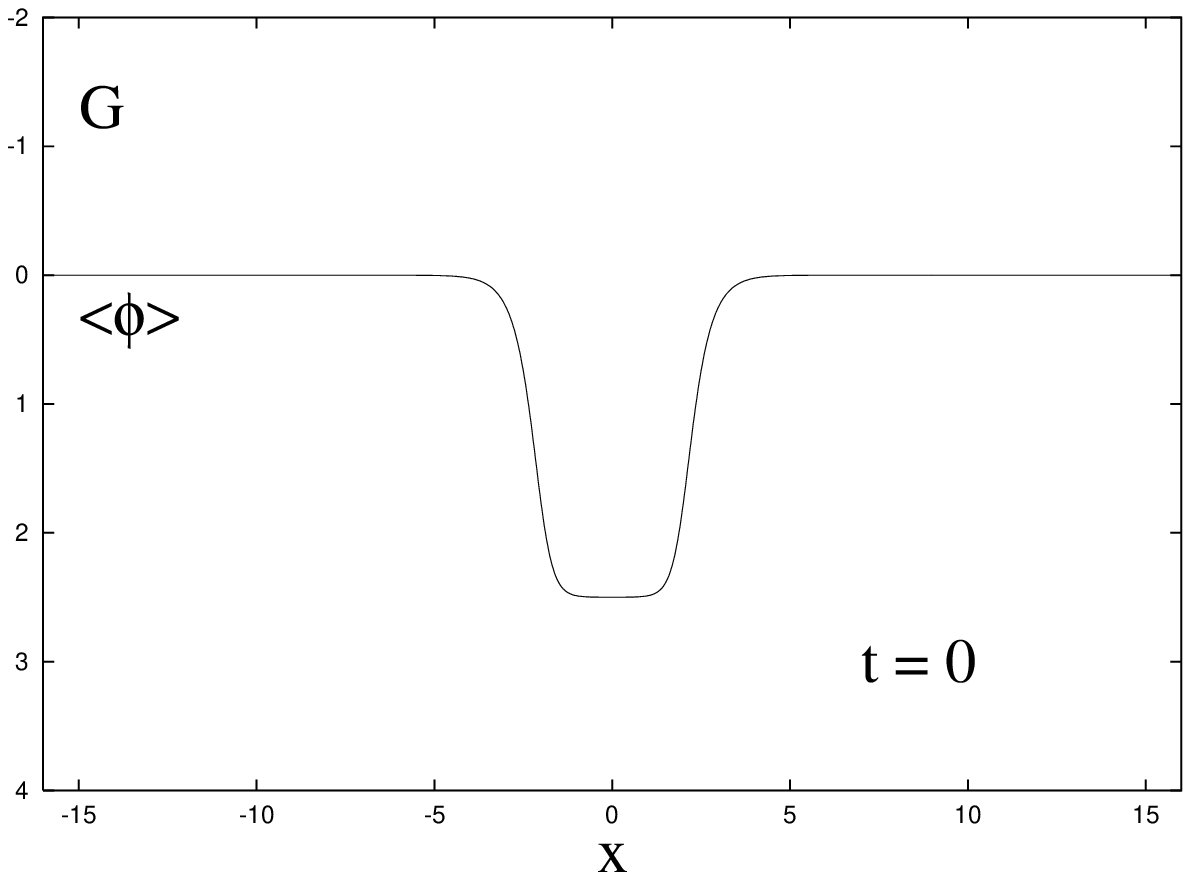,width=1.65in,height=1.12in,silent=,angle=0}
\psfig{file=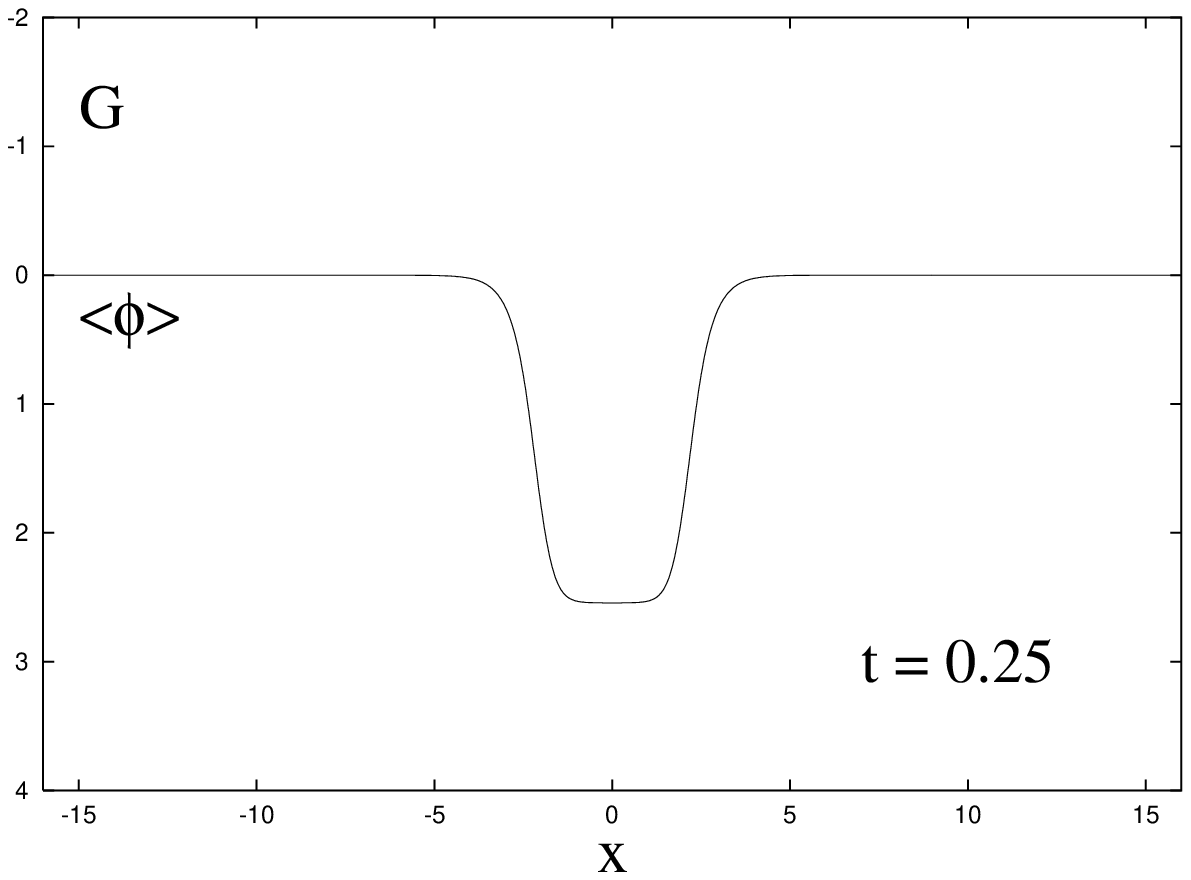,width=1.65in,height=1.12in,silent=,angle=0}
\psfig{file=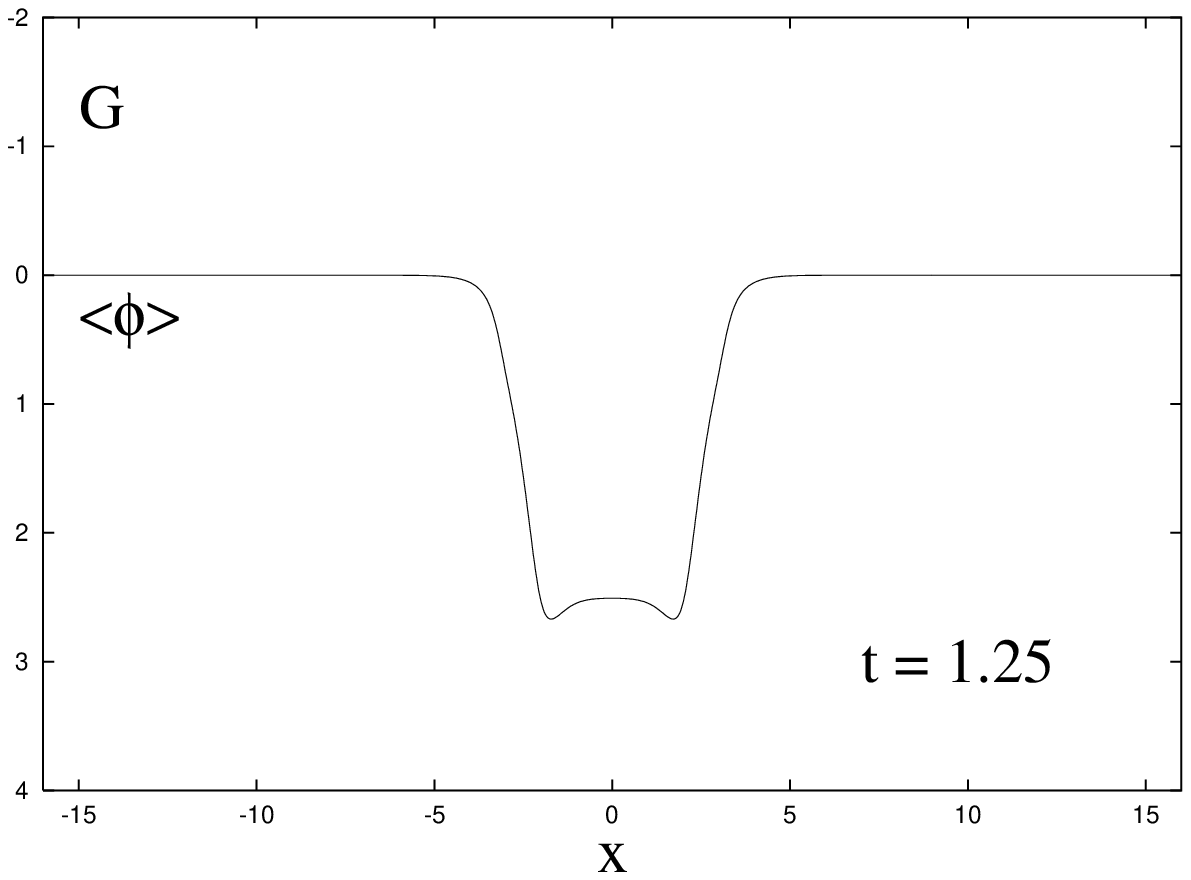,width=1.65in,height=1.12in,silent=,angle=0}
\psfig{file=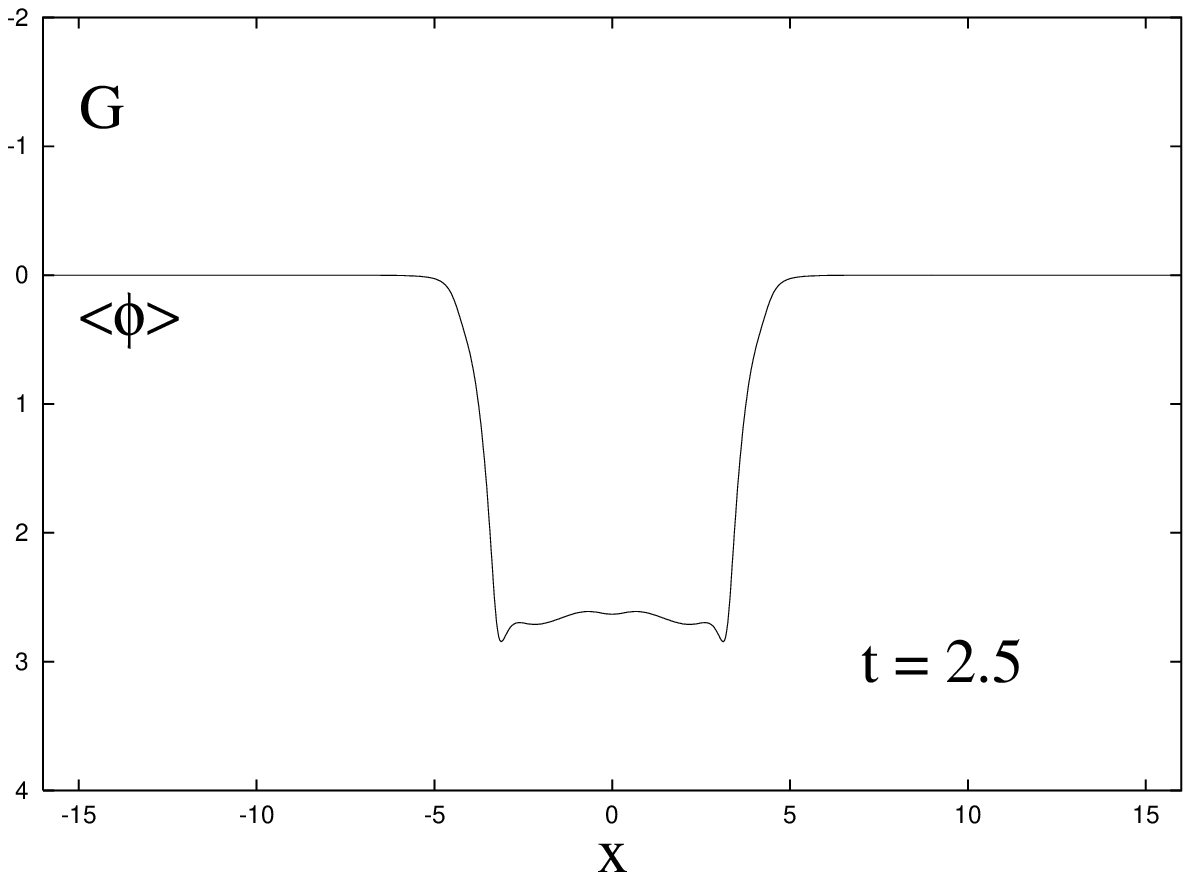,width=1.65in,height=1.12in,silent=,angle=0}
\end{center}
\caption{Snapshots of the expectation value for classical and
quantum field (lower curves) and two-point function $G$ (upper curve)
at increasing time intervals. The initial
configuration shown in the top left is almost identical in both cases.
The evolution shows the quantum bubble relaxing into the minimum
energy configuration preferred by the quantum effective potential while
being screened by quantum fluctuations. 
In so doing, it leaves behind excitations inside the bubble and
accelerates more rapidly. For presentation, $G$ is shown
shifted away from its actual value (equal to zero where $\phi=0$.)}
\label{snapshots}
\end{figure}

The first clear difference is that the field profile corresponding 
to the bubble is no longer purely classical.  The quantum 
modes respond to our (initially) classical background and partially 
screen it. The result is a dressed bubble, an object with 
a quasi-classical profile around which the vacuum is disturbed. This 
self-consistent object is Lorentz invariant and once formed propagates 
adiabatically, i.e. without further particle creation. Several 
snapshots  of the mean field and 2-point function in Fig.~\ref{snapshots} 
illustrate this behavior, from initial formation to late propagation.  

The next important effect of quantum fluctuations is essentially static. 
Looking at Fig.~\ref{snapshots} we also note that while the
``height'' of the dressed bubble (i.e. the true vacuum expectation value (vev)
of the field) is initialized at the classical vev, it will not remain
at that value during bubble growth.  This is nothing remarkable, only
a demonstration of the well known fact that the quantum effective
potential differs from the classical potential, and their minima do not coincide.  It
might make more sense to initialize the field with the true vacuum
expectation value given by the quantum effective potential at some
loop order.  However, given the naive initialization of the two-point function that we employ
here, the field will still be perturbed away from this minimum inside
the bubble when the two-point function changes.  The analyticly computated quantum effective
potential treats the mean field as spatially
homogeneous.  Therefore it cannot be expected to describe accurately the dynamics
of the bubble.  We discuss this
problem and its solution in more detail below and in Section V.

\begin{figure}
\begin{center}
\leavevmode
\psfig{file=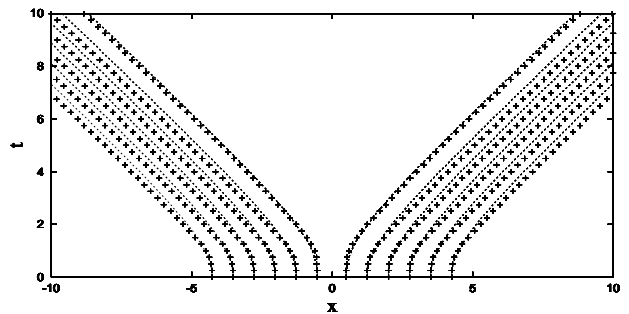,width=3.2in,height=1.6in,silent=,angle=0}
\end{center}
\caption{The quantum Hartree approximation trajectories for initial
bubbles of various sizes are shown as pluses.  The lines are
interpolations of the classical trajectories shown in
Fig.~\ref{trajectories}.  The initial acceleration of the quantum
bubble wall is manifestly more rapid, and it approaches its lightcone asymptote
 more quickly.}
\label{qmtrajectories}
\end{figure}

At zero temperature, we observe both super- and
sub-critical bubbles in the quantum case which grow 
and decay respectively as in the classical case.  We track the
position of the bubble wall in time.  Despite the fact that the
quantum evolution introduces fluctuations absent in the
classical case, we see a bubble wall trajectory which is consistent
with the classical picture.  We observe approximately hyperbolic 
trajectories which asymptotically approach the speed of light, as
shown in Fig.~\ref{qmtrajectories}. Fig.~\ref{snapshots} shows some of the
quantitative differences between the classical and the quantum
evolution, including the effect of the ``roll'' down to the minimum of
the quantum effective potential.  As the field vev changes inside the
bubble, it releases potential energy beyond what would happen
from the movement of the interface--namely bubble growth--alone.  Since
this energy is converted into kinetic and gradient energy of the field
and fluctuations, this roll can be held responsible for a more rapid
acceleration at the initial stage of growth.  The energy of the true
vacuum in the quantum case is also lower, so the rate of energy
release by the bubble growth is larger.  Hence, we see the quantum
trajectories in Fig.~\ref{qmtrajectories} approach their asymptotic
value (again unity) more quickly.  This is the only significant departure from
the particular hyperbolic trajectory predicted by a classical analysis.

We also note that while the classical critical radius for growth
was found to be $R_c=0.35$, the critical radius in the quantum evolution 
is around $R_c=0.45$. 
In both cases, the critical radius for growth
is a factor of 4 to 5 smaller than the bounce radius $R_B$ while
relatively close (within a factor of 2) to the lower limit in
(1+1)-dimensions for a wall with some thickness $\mu^{-1}$ (i.e. $R_c
\simeq 1/\mu$).  In Section~\ref{secII}, we found $\mu^{-1}=0.24$.

It is worth considering whether the accelerated initial growth and the
observed change in $R_c$ can be understood just from a classical
analysis of the bubble using the effective potential instead of the
bare potential.  Recall that the particular shape of the hyperbolic
trajectory is determined in the  thin-wall analysis by
the shape of the potential through the parameter $R_B$. For example,
we can compute the effective potential 
numerically using homogeneous fields and solving a gap equation along
the lines of earlier works \cite{DJ}.  From this, we can extract a quasi-classical
potential and recalculate $R_B$ and $\mu^{-1}$ for this potential.  We
obtain $R_B = 0.86$ and $\mu^{-1}=0.25$.   We fit the observed value
of $R_B=1.3$ from the trajectories in Fig.~\ref{qmtrajectories}.
While the classical thin-wall predictions using the quantum
effective potential give the correct qualitative changes, 
they predict the wrong magnitude of the correction.  
We should also note that parametrically the thin wall approximation ought to 
be worse in the quantum case since now $\mu R \simeq 3.44$, less than  
half its classical value.

This rough calculation harkens back to several efforts at applying the
bounce analysis to quantum field models with, for example, symmetry 
breaking due to radiative corrections~\cite{Weinberg}.  
The problem in effect stems from the fact that the degrees of freedom 
which need to be traced over in the calculation of the effective 
potential cannot be properly integrated out since they participate in 
the bubble dynamics.
We believe that a way out of this dilemma necessitates calculation of
a self-consistent bounce, i.e. not only the shape of the mean field
but also the full spectrum of interacting fluctuations in its background 
as prescribed by an effective action.  
While this is beyond the scope of the present work, we are pursuing it for 
future publication. Furthermore, with this information in hand, one 
could hopefully also bring to bear an analysis of the dynamical viscosity 
experienced by the moving wall at finite temperature, as suggested 
by other recent results~\cite{CalHuRamsey,Boyanovsky}.

In the meantime, the observed differences we have described in the 
shape and trajectory of the purely classical bubble and its dressed 
quantum counterpart are all that one may expect.  
At zero temperature, the qualitative
characteristics of bubble propagation remain determined by the
constraint of Lorentz invariance.  Once finite temperature is considered,
or alternately in the presence of a Lorentz-violating condensate, bubble propagation
can change drastically, and the velocity of domain growth will in
general not asymptote to that of light.  

Turning now to the last of Coleman's open questions, we consider the
effect of colliding bubbles in the classical and quantum Hartree
approximations.  Due to the nature of our numerical simulations, we
enforce periodic boundary conditions on our one-dimensional lattice at
the endpoints.  When the bubble interface reaches the end of the
lattice, it effectively meets its mirror image or, equivalently, a neighboring bubble wall.  Whether
the interfaces bounce off of each other or coalesce can then be
observed.  

\begin{figure}
\begin{center}
\psfig{file=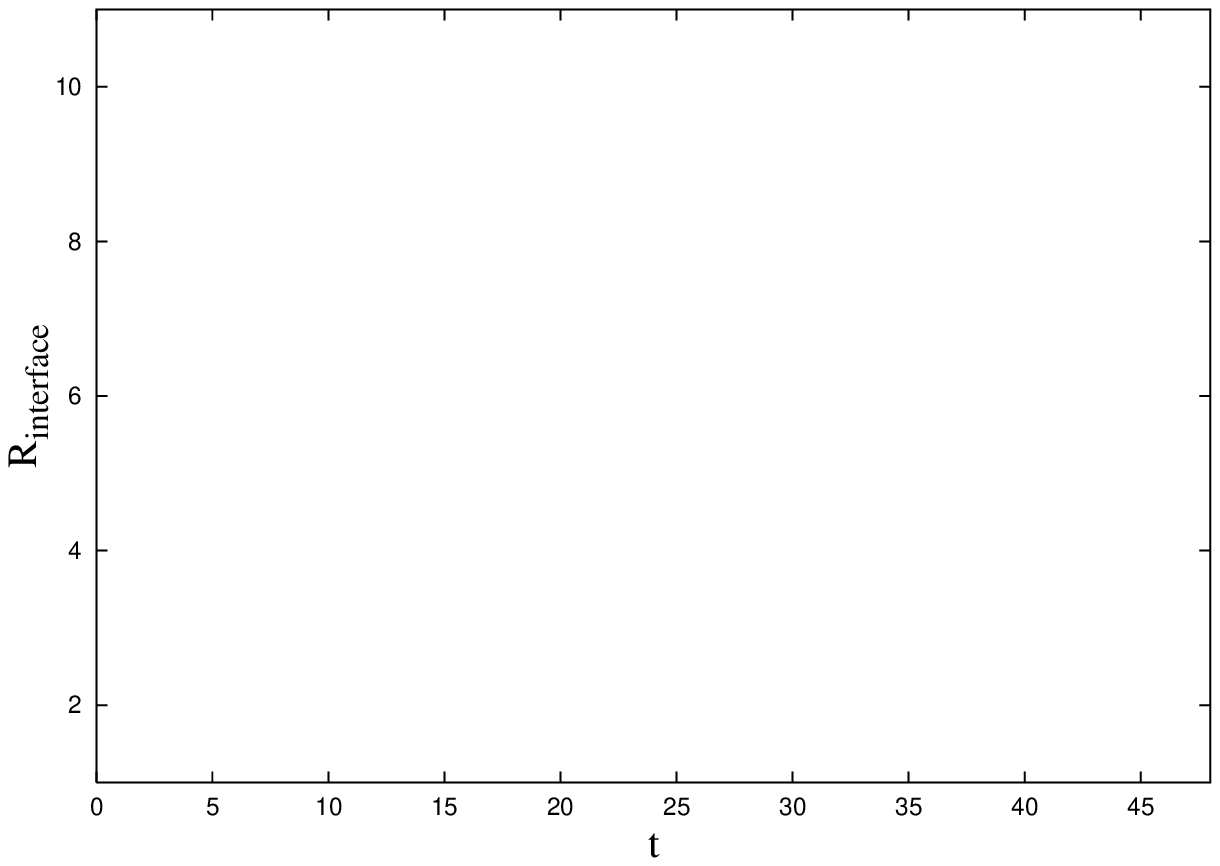,width=3in,height=2in,silent=,angle=0}
\end{center}
\caption{Trajectories for bubble interfaces bouncing off of
neighboring interfaces. Connected curves show classical trajectories for
the bounce (bottom) and distortions thereof (upper curves), while quantum trajectories are shown
as data points.  The classical bounce and the quantum data are plotted for
lattice spacings of 1/64 and 1/128, but the trajectories overlap
entirely.  The quantum trajectory loses coherence quickly relative
even to the distorted classical bubbles.} 
\label{bubblecollisions}
\end{figure}

We note a surprising result: classically the bubble walls do not coalesce.
The collision of the interfaces does temporarily excite
wavelengths near the bubble ``penetration depth'' (the bubble thickness) at low amplitude, but this energy
appears to be re-exchanged with the bubble interface.  Furthermore, this is a
special feature of the bounce profile which is not robust under deformations
of the shape of the bubble:  other initial bubbles do radiate and
decay albeit slowly through collisions, see
Fig.~\ref{bubblecollisions}.  We also verified that the stability of the bounce disappears in
(2+1)-dimensions where collisions rapidly destroy the spherical
symmetry of the bubble.  

The (1+1)-dimensional classical bounce
solution exhibits truly solitonic behavior in the sense that it is
unaltered by scattering off of another (solitonic) bubble.  This
effect, well known to exist in integrable nonlinear
systems~\cite{solitons}, is unexpected in our model.
In fact, Lohe has considered the linearized
perturbations around the static soliton (\ref{thinbubble}) of
this model as one example in a parameter family of polynomial
potentials (the sine-Gordon model appears as a limiting case); he
has shown that there are completely reflected states in one direction
from the soliton or bubble wall~\cite{Lohe}.
This is just a physical consequence of the difference in the
mass of elementary excitations about the two vacua: low-lying
states in the false vacuum cannot pass through the interface because
they have no energy counterpart on the other side.  We may take
this as suggestive evidence that the two bubble interfaces must repel
under these circumstances, although it cannot guarantee the observed
almost perfect reflection.  If the sign of the $\gamma$-term in
the potential~(\ref{potential2}) is flipped, the initial bubble will
collapse and the interfaces will pass through each other, exploiting
the $\phi \rightarrow -\phi$ symmetry of the model as they emerge on
the other side of the collision region..  We verified this behavior
numerically.
The classical scattering of our interfaces is thus quite analogous to the
scattering of solitons in the sine-Gordon model where both types of
behavior (perfect reflection and perfect transmission) obtain even though
the vacua are identical~\cite{SG}. In that case, it is soliton-soliton
or soliton-antisoliton solutions which display the two distinct
possibilities.

Quantum bubbles by contrast appear to dissipate energy rather
efficiently during collisions and especially during the propagation of the
interface through the fluctuations created  by the collision.
Although at the level of our approximation the bubbles still rebound
initially from the collision,  they quickly lose kinetic energy.
By the second or third collision, all semblance of the initial
configuration is lost. In order to understand the quantum, or at least
semi-classical scattering of the interfaces, one could  perform a
detailed study along the lines of Ref.~\cite{RJWOO}, adapted for the
self-consistent quantum evolution.

\begin{figure}
\begin{center}
\psfig{file=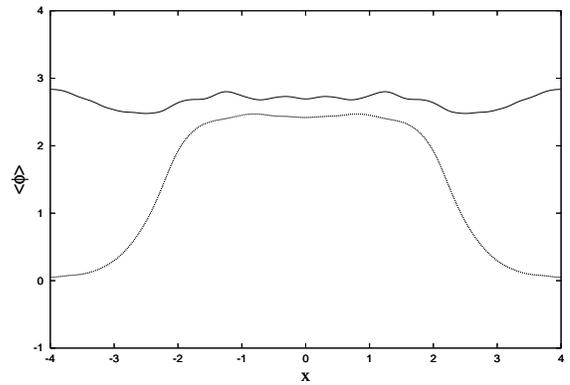,width=3in,height=2in,silent=,angle=0}
\end{center}
\caption{Field expectation values for the quantum (upper) and classical (lower) evolution after twelve collisions with neighboring bubbles (effected by means of periodic boundary conditions).  The classical bubble still maintains its shape, while the quantum bubble appears to have dissipated its energy into fluctuations around the true vacuum.} 
\label{collisionsnapshot}
\end{figure}

We present results in Fig.~\ref{bubblecollisions}, where we plot the
classical and quantum trajectories at two different lattice spacings
(same physical volume) in order to demonstrate that this effect is not an artifact of the
lattice discretization.  We also plot classical trajectories of
distorted bubbles to show that they decay on a much longer time scale
than the quantum bubbles. The trajectories are obtained by following
one field value near the true vacuum; in the quantum and classical distorted cases, the
apparent loss of periodicity indicates that fluctuations around the
true vacuum no longer have a coherent structure.  In Fig.~\ref{collisionsnapshot} we plot a
late time snapshot of the field evolution in the classical and quantum cases.

\section{Discussion and conclusions}
\label{secV}

In this paper we took the first steps towards showing how self-consistent 
quantum fluctuations can be incorporated in the real time dynamics 
of bubble interfaces. We have shown that a classical bubble profile
becomes dressed by quantum fluctuations, which in turn affect the rate 
of conversion of the false to true vacuum, i.e. the acceleration of the 
bubble velocity towards the speed of light.  We have also found that
the presence of quantum fluctuations promotes substantially more
efficient transport of the bubble wall energy into particles at bubble 
collisions.

There is one inelegant feature of our analysis, namely that the initial
conditions for the quantum evolution were informed by the classical
bounce calculation in absentia of self-consistent fluctuations.  There
should exist a solution to the coupled Dyson-Schwinger equations which
is analogous to the bounce, the analytic continuation of which amounts
to a dressed bubble which expands without radiation (particle creation).
Using this configuration as a starting point for an initial value
problem would eliminate the unwanted radiation that we observed
resulting from the early time response of the vacuum to the nontrivial
classical background.  

Starting from the 2PI-CTP formalism it is natural to generalize
the one-loop computation of the false vacuum decay rate to a self-consistent problem, where fluctuations 
and the mean-field bubble profile are solved together.  The
self-consistent nucleation rate and bubble evolution would be accurate to the
same level of truncation as the effective action for any approximation,
not limited to the Hartree example above.  We believe that this
intermediate step is necessary before more subtle features of the
long time evolution--inevitably entangled with the early time
response--can be analyzed quantitatively in detail.  We 
have made progress towards this implementation, which we intend to
report in a future work~\cite{nextpaper}.

Knowledge of the spectrum of self-consistent fluctuations will
render the inclusion of medium effects such as finite temperature more realistic.
It also paves the way for other generalizations of the model and
improvements upon the approximation we have employed.  A finite
temperature analysis in particular may be sensitive to higher loop
order interactions which give rise to viscosity effects in linear
response around equilibrium.  Through analysis of the energy-momentum
tensor $\vev{T_{\mu\nu}}$ we can
investigate the emergence of hydrodynamic interface propagation
from microscopic physics.  We also intend to extend the
model to include fermions. 

There are many more directions in which a similar analysis may be
informative.  From numerical solution of a subcritical bubble which
decays, we can read the spectrum of asymptotic states.  It may be
possible that the same states, upon time reversal, would provide the initial
conditions necessary to generate a bubble.  Such an asymptotic configuration
is generically a complicated, correlated many-particle state; however its overlap
with a two-body state may hint at the probability of a bubble resonance
in a scattering experiment.  Stable dynamical solutions such as
breathers have long been features of classical field theories.  It is
unknown whether they are stable upon inclusion of dynamical
quantum effects.  Finally, there has been a great deal of excitement
recently about extended objects such as branes and their dynamics,
collisions etc.  If there are fields confined to each brane in a
homogeneous manner, then the effect of colliding two branes can be
cast  into a problem very similar to the one explored here with the
addition of traces over transverse degrees of freedom.

\begin{acknowledgments}
We are grateful to K. Rajagopal for many useful discussions and suggestions 
and for comments on the manuscript. We also acknowledge helpful
comments from J. Berges, R. Jackiw, L. Levitov and S. Todadri.
This work was supported in part by the D.O.E. under research agreement 
$\#$DF-FC02-94ER40818.  Y.~B. is also supported by an NSF Graduate Research Fellowship. 
\end{acknowledgments}

\end{document}